\documentclass[12pt,useAMS,usenatbib]{mn2e}
\usepackage{multirow}
\usepackage{graphicx}
\usepackage{multicol}
\usepackage{caption}
\usepackage{subcaption}

\defcitealias{mtb12b}{MTB12a}

\title[Monitoring the Galactic Centre with Australia Telescope Compact Array]{Monitoring the Galactic Centre with Australia Telescope Compact Array}
\author[Borkar et. al.]{Borkar, A.$^{1,2}$\thanks{E-mail:borkar@ph1.uni-koeln.de;},
Eckart, A.$^{1,2}$, Straubmeier, C.$^{1}$, Kunneriath, D.$^{3}$, Jalali, B.$^{1}$, 
\newauthor  Sabha, N.$^{1}$, Shahzamanian, B.$^{1,2}$, Garc\'{i}a-Mar\'{i}n, M.$^1$, Valencia-S, M.$^{1}$, 
\newauthor Sjouwerman, L.$^{4}$, Britzen, S.$^{2}$, Karas, V.$^3$, Dov\v{c}iak, M.$^3$, Donea, A.$^{5}$, Zensus, A.$^{2,1}$
\vspace{0.4cm}\\
$^{1}$ I. Physikalisches Institut, Universit\"at zu K\"oln, Z\"ulpicher Strasse 77, D-50937, K\"oln, Germany\\
$^{2}$ Max-Planck-Institute f\"ur Radioastronomie, Auf Dem H\"ugel 69, D-53121, Bonn, Germany\\
$^{3}$ Astronomical Institute of the Academy of Sciences, Prague, Bo\v{c}n\'{i} II 1401/1a, CZ-14131 Praha 4, Czech Republic\\
$^{4}$ National Radio Astronomy Observatory, PO Box 0, Socorro, NM 87801, USA\\
$^{5}$ Monash Centre for Astrophysics, Monash University, Clayton, Victoria 3800, Australia}

\begin{document}

\date{Accepted 2016 February 12. Received 2016 February 12; in original form 2015 May 21}

\pagerange{\pageref{firstpage}--\pageref{lastpage}} \pubyear{2016}

\maketitle

\label{firstpage}

\begin{abstract}

The supermassive black hole, Sagittarius A* (Sgr A*), at the centre of the Milky Way undergoes regular flaring activity which is thought to arise from the innermost region of the accretion flow. We performed the monitoring observations of the Galactic Centre to study the flux-density variations at 3mm using the Australia Telescope Compact Array (ATCA) between 2010 and 2014. We obtain the light curves of Sgr A* by subtracting the contributions from the extended emission around it, and the elevation and time dependent gains of the telescope. We perform structure function analysis and the Bayesian blocks representation to detect flare events. The observations detect six instances of significant variability in the flux density of Sgr A* in three observations, with variations between 0.5 to 1.0 Jy, which last for $1.5-3$ hours. We use the adiabatically expanding plasmon model to explain the short time-scale variations in the flux density. We derive the physical quantities of the modelled flare emission, such as the source expansion speed $v_{\mathrm{exp}}$, source sizes, spectral indices, and the turnover frequency. These parameters imply that the expanding source components are either confined to the immediate vicinity of Sgr A* by contributing to the corona or the disc, or have a bulk motion greater than $v_{\mathrm{exp}}$. No exceptional flux density variation on short flare time-scales was observed during the approach and the flyby of the dusty S-cluster object (DSO/G2). This is consistent with its compactness and the absence of a large bow shock.

\end{abstract}


\section{Introduction}

It has been established that the centre of the Milky Way harbours a supermassive black hole (SMBH) with a mass of $4 \times 10^6 M_\odot$, which is associated with the compact radio source, named Sagittarius A* (Sgr A*) at a distance of $8$ $kpc$ from Earth (\citealt{Eckart-Genzel96, Genzel97,Genzel00, Eckart02, Schodel02, Schodel03, Eisenhauer03, Eisenhauer05, Ghez04, Ghez05, Ghez09, Gillessen09a, Gillessen09b}). Multiwavelength observations at radio, submillimeter, IR and X-ray wavelengths have shown that the bolometric luminosity of Sgr A* is several orders of magnitude less than its corresponding Eddington luminosity. Several theoretical models have been proposed to explain the faint emission, which require a very inefficient radiation mechanism for the predicted mass inflow of the accreting gas (\citealt{Narayan98, Liu-Melia01, Melia-Falcke01, Yuan03ApJ, Goldston05, Liu06a}). A fairly successful modelling of the multiwavelength spectrum of Sgr A* was done using the advection dominated accretion flow (ADAF) model (\citealt{Narayan98}), which predicts a two-temperature radiatively inefficient flow which is advected across the event horizon. Estimates from the observations of radio and submillimeter polarisation (\citealt{Bower03, Marrone06}) predict significantly lower electron densities in the accretion region of Sgr A*, implying that most of the material does not reach the central black hole. The study of the flare emission from Sgr A* allows us to contribute to the solution of this issue.

Studies of the variability of Sgr A* have presented the intra-day flux density excursions (called `flares') of Sgr A* from radio to submillimeter wavelengths (\citealt{Bower02, Zhao03, Herrn04, Mauerhan05, Li09}). These flares have been demonstrated to follow the brightest simultaneous NIR/X-ray flares with a delay of $\sim$ 100 min (\citealt{Eckart04, Eckart06a, Eckart08b, Eckart12, Marrone08, YZ06b, YZ08}). The synchrotron self Compton (SSC) model has been used to explain the flare emission at these wavelengths. The model involves up-scattering of the sub-mm photons from a compact source component. The inverse Compton scattering of the THz-peaked flare spectrum by relativistic electrons gives rise to the X-ray flares while the NIR flares arise due to the synchrotron and SSC mechanism \citep{Markoff01, Liu-Melia02, Yuan03ApJ}. The flares in radio and submillimeter regime are caused by the adiabatic expansion of the source components \citep{Eckart08b, Eckart12}. The study of the flaring activity gives us indirect constraints on the details of the emission mechanism and the geometry of the source.

In this paper, we report the results of our observations of the GC at 3.5 mm, taken between 2010 to 2014 using the ATCA.\@ In Section 2, we describe the observations and data reduction processes used to obtain the light curves for Sgr A*. The results of the observations are discussed in Section 3, where we report the detection of significant variability seen in the light curve in 3 epochs using structure function analysis and Bayesian blocks representation. The modelling and analysis of the observed flares using the adiabatic expansion model are discussed in Section 4, followed by summary and conclusions in Section 5.

\section[]{Observations and Data Analysis} 


Between 2010 and 2014, we observed the GC at 3 mm wavelength with the ATCA.\@ ATCA is an array of six 22-m telescopes located at the Paul Wild Observatory in Narrabri, NSW, Australia, of which five are available for 3mm observations. The location of ATCA allows us to observe the GC for more than 8 hours a day, as the GC passes almost overhead in the southern hemisphere. The Compact Array Broadband Backend (CABB), upgraded in 2007, allows for observations with two 2048 MHz intermediate frequency (IF) bands. We performed the observations with the spectral line mode wherein we observed at two different frequency bands at 86.243 and 85.640 GHz, corresponding to two transition lines of the SiO maser (SiO $J = 2-1$, $v = 1$ and $J = 2-1$, $v = 2$). The GC was observed for approximately $10-12$ hours per observation day with bandpass calibration for 30 min with $\mathrm{PKS}$ $1253-055$ before the observations and flux calibration with Uranus at the end of the observations. Sgr A* was observed with three sets of 25 min on-source observations which are in-between the observations of gain calibrators (see Table~\ref{Table1}). Sgr A* is strong enough that it can be used as a self calibrator for phase calibrations. The details of the observations are summarised in Table 1. The interferometer data was mapped and reduced using the MIRIAD data reduction package.

\begin{table*}
 \centering
 \begin{minipage}{140mm}
  \caption{The Log of observations of Sgr A* taken during 2010 to 2014. The dashes represent the days on which observations were not made. See section~\ref{Results} for details.}\label{Table1}
  \begin{tabular}{c c c c l l}
  \hline\hline

   \multicolumn{2}{c}{Date} & Array & Calibrators & UT and JD & UT and JD\\
   & & & & Start Time & End Time\\
    \hline
   \multirow{8}{*}{2010} & May 13 & \multirow{8}{*}{H214} & 1741- 312 & 2010 May 13 11:04:45 & 2010 May 13 21:50:25\\ 
   & & & & JD2455329.96042 & JD2455330.41001\\
   & May 14 & & 1622-297 & 2010 May 14 10:45:07 & 2010 May 14 22:07:40\\ 
   & & & & JD2455330.948 & JD2455331.42199\\
   & May 15 & & 1741-312 & 2010 May 15 10:10:30 & 2010 May 15 22:30:10\\ 
   & & & & JD2455331.92406 & JD2455332.42406\\
   & May 16 & & 1622-297 & 2010 May 16 10:08:47 & 2010 May 16\\ 
   & & & & JD2455332.92277 & JD2455333.41961\\
   \hline
   \multirow{8}{*}{2011} & May 23 & \multirow{8}{*}{H214} & 1741-312 & 2011 May 23 09:57:43 & 2011 May 23 21:05:13\\ 
   & & & & JD2455704.91516 & JD2455705.37862\\
   & May 24 & & 1741-312 & 2011 May 24 09:56:05 & 2011 May 24 21:24:17\\ 
   & & & & JD2455705.91395 & JD2455706.39186\\
   & May 25 & & 1741-312 & 2011 May 25 10:01:13 & 2011 May 25 21:22:30\\ 
   & & & & JD2455706.91751 & JD2455707.39062\\
   & May 26 & & 1741-312 & 2011 May 26 10:03:01 & 2011 May 26 21:17:33\\ 
   & & & & JD2455707.91876 & JD2455708.38719\\
   \hline
   \multirow{8}{*}{2012} & May 15 & \multirow{8}{*}{H214} & 1714-336 & 2012 May 15 08:23:47 & 2012 May 15 21:51:35\\ 
   & & & & JD2456062.84985 & J2456063.41082D \\
   & May 16 & & 1714-336 & 2012 May 16 10:04:30 & 2012 May 16 21:45:22\\ 
   & & & & JD2456063.91979 & J2456064.4065D\\
   & May 17 & & 1714-336 & 2012 May 17 09:49:17 & 2012 May 17 21:52:18\\ 
   & & & & JD2456064.90922 & JD2456065.41132\\
   & May 18 & & 1714-336 & 2012 May 18 11:02:51 & 2012 May 18 21:56:46\\ 
   & & & & JD2456065.96031 & JD2456066.41442\\
   \hline
   \multirow{12}{*}{2013} & June 26 & EW352 & 1741-312 & 2013 June 26 08:18:21 & 2013 June 26 20:44:37\\ 
   & & & & JD2456469.84608 & JD2456470.36432\\
   & June 27 & EW 352 & 1741-312 & 2013 June 27 --- & 2013 June 27 ---\\ 
   & & & & --- & ---\\
   & August 31 & 1.5A & 1741-312 & 2013 August 31 03:36:08  & 2013 August 31 14:57:04 \\ 
   & & & & JD2456535.65009 & JD2456536.12296\\
   & September 1 & 1.5A & 1741-312 & 2010 September 2 --- & 2013 September 2 --- \\ 
   & & & & --- & ---\\
   & September 14 & H214 & 1741-312 & 2013 September 14 03:08:12 & 2013 September 14 13:27:58\\ 
   & & & & JD2456549.63069 & JD2456550.06109\\
   & September 16 & H214 & 1741-312 & 2013 September 16 --- & 2013 September 16 --- \\ 
   & & & & --- & ---\\
   \hline
   \multirow{10}{*}{2014} & April 1 & H168 & 1741-312 & 2014 April 1 14:54:24 & 2014 April 1 23:38:55\\ 
   & & & & JD2456749.12111 & JD2456749.48536\\
   & April 2  & H168  & 1741-312 & 2014 April 2 13:37:58 & 2014 April 3 00:22:18 \\ 
   & & & & JD2456750.06803 & JD2456750.51549\\
   & June 7 & EW352 & 1714-336 & 2014 June 7 08:00:50 & 2014 June 7 19:00:02\\ 
   & & & & JD2456815.83391 & JD2456816.29169\\
   & September 26 & H214 & 1714-336 & 2014 September 26 02:05:40 & 2014 September 26 12:44:31\\ 
   & & & & JD2456926.58727 & JD2456927.03091\\
   & September 27 & H214 & 1714-336 & 2014 September 27 02:10:13 & 2014 September 27 12:40:51\\ 
   & & & & JD2456927.59043 & JD2456928.02837\\

\hline

\end{tabular}

\end{minipage}
\end{table*}

The antenna gain of ATCA depends significantly on the elevation angle and has a maximum efficiency at the elevation angle of $60^{\circ}$ and a minimum value at the elevation angle of $90^{\circ}$. Further, the gravitational distortion of the dish limits the antenna efficiency at angles below $40^{\circ}$. The antenna gain is also dependent on the angular distance between Sgr A* and the calibrator (\citealt{Li09}). The gain corrections derived from a distant calibrator cannot fully compensate for the elevation effects, which will show up in the light curve. To avoid this, we used nearby calibrators within $10^{\circ}$ of Sgr A* as the gain calibrators, mainly $\mathrm{PKS}$ $1741-312$ which is only $2.29^{\circ}$ away. We also used secondary gain calibrators to check for consistency. At millimeter wavelength, the atmosphere is no longer transparent and can cause large variations in the gain. Strong variation due to the thick atmosphere is observed at lower elevation angles (less than $45^{\circ}$). Also, for some array configurations, shadowing is significant for elevations of $<40^{\circ}$.  Thus we only use the observational data above the elevation of $40^{\circ}$. 

The unavailability of the 6th antenna for the 3mm observations means that the maximum baseline available is 214 m for the H214 configuration which we have used predominantly. This gives a synthesized beam of 2\arcsec. 
At this angular resolution the diffuse emission from the extended sources surrounding Sgr A*, such as the mini-spiral, dominates the flux density at shorter baselines. This emission also contributes to the flux at longer baselines. Table~\ref{minflux} shows the median flux density of Sgr A* obtained from the light curves for the observations. The median value is assumed to represent the observed flux density value around which flux variations are detected. These values are consistent with the flux density values from the ATCA database, and also compatible with the recent VLA observations of Sgr A* (\citealt{Sjou14}). The flux contains contribution from the extended emission around Sgr A*. A 500 minute ($\sim 8$ hour) variability in the flux density of the Sgr A* has been indicated by \citet{Witzel12} and \citet{Dex14}, which may account for the difference in the day to day flux density values.

\begin{table}
 \centering
 \begin{minipage}{70mm}
  \caption{The median flux density values of the Sgr A* for the observations.}
   \begin{tabular}{l c}
     \hline
     Observation Day & Flux density [Jy]\\
     \hline
     13 May 2010 & 4.97\\
     14 May 2010 & 4.32\\
     15 May 2010 & 4.66\\
     16 May 2010 & 3.99\\
     & \\
     23 May 2011 & 3.93\\
     24 May 2011 & 3.63\\
     25 May 2011 & 3.35\\
     26 May 2011 & 3.56\\
     & \\
     15 May 2012 & 3.80\\
     16 May 2012 & 3.67\\
     17 May 2012 & 3.22\\
     18 May 2012 & 3.26\\
     & \\
     26 June 2013 & 4.16\\
     31 August 2013 & 4.38\\
     14 September 2013 & 6.73\\
     & \\
     01 April 2014 & 2.68\\
     02 April 2014 & 2.80\\
     07 June 2014 & 2.94\\
     26 September 2014 & 3.32\\
     27 September 2014 & 3.99\\
\hline
   \end{tabular}

\label{minflux}
\end{minipage}
\end{table}

These contributions need to be accounted for to obtain the flux density intrinsic to Sgr A*. To correct for the extended flux contributions as well as for the atmospheric and instrumental variations in the flux of Sgr A*, we used the method developed by \citet{Kunneriath10}. First the flux values from two orthogonal pairs of the longest baselines were extracted, and the median baseline and time dependent flux was subtracted for each baseline from each dataset of observations. For each baseline $b$, $s_b(t)$ represents the flux density observed for that baseline. Then $S_b(t)$ gives the median baseline flux for each year
\begin{equation}
S_b(t) = \mu_{yr}(s_b(t)).
\end{equation}

The time dependent differential flux density $I(t)$ and the uncertainties $\delta I(t)$ were calculated using

\begin{equation}
I(t) = \mu_b(s_b(t) - S_b(t))
\end{equation}

and

\begin{equation}
\delta I(t) = \mu_b(S(t) - (s_b(t) - S_b(t)))
\end{equation}

Here $\mu_{yr}$ represents the median operator with the median taken over the observation epochs for each year, while $\mu_b$ is the median calculated over different baseline. The uncertainty in the differential flux density $\delta I(t)$ is the median of the deviation from the median flux $I(t)$. As the differential flux takes both positive and negative values, a constant flux was added or subtracted to the differential flux densities to model the flux densities. We attribute the residual flux density dips/excesses to variations in the intrinsic flux density of Sgr A*.

\section[]{Results}\label{Results}

The observations at 3mm with ATCA are very sensitive to atmospheric conditions. On several observing days, the light curves are contaminated by atmospheric turbulence or rainfalls, which manifest as very strong and rapid variations at very short timescales ($\sim 30$ min). In 2010, turbulence affected most of the observation on 16 May. On 23 May 2011, a heavy rainfall caused extreme variations in the differential flux density as large as 7 Jy. Similar rapid variations were observed on 26 May 2011 and 15 May 2012, where the turbulence affected significant portions of the observations.

In 2013, heavy rainfall on three days prevented from observing the GC and no data was obtained on 27 June, 1 September and 16 September (marked by `$-$' in the Table 1). On 31 August and 14 September, large fluctuations in part of the data were observed due to turbulence while during the remaining time, the flux remained fairly stable. No significant fluctuations were observed on 26 June 2013 either.

The observations in 2014 were not affected by any atmospheric events. The conditions were good and we obtained good quality flux data for all observation days. Although the data quality was good, we do not detect any flaring activity on any day of 2014 of the time-scale of $\sim 150$ minutes within the observation times.

\subsection{Structure Function Analysis}

The structure function analysis has been extensively used to study the light curves and to characterise the variability in AGN light curves (\citealt{Simonetti85, Hughes92, Paltani99}), and in Sgr A* light curves (\citealt{Do09, YZ11, Dexter14}). The first order structure function is defined as

\begin{equation}
SF(\tau) = \langle {[I(t)-I(t+\tau)]}^2 \rangle
\end{equation}

where $I(t)$ is the signal at time $t$ and $\tau$ is the time lag. For an unevenly sampled data, the structure function is calculated by measuring the ${[I(t)-I(t+\tau)]}^2$ for all possible values of lag $\tau$ and placing them into bins. The value of the structure function for a bin is the average of all values of SF in that bin, while the median time lag of each bin is assumed as the lag for that bin. The errors are calculated using $\sigma_{bin}/\sqrt{N_{bin}}$ where $\sigma_{bin}$ is the standard deviation of the data points in each bin and $N_{bin}$ is the number of points in each bin.

Ideally, the structure function shows two plateaus at the two ends of the curves which are connected by a power law portion. The plateau at the lower lag values is proportional to the measurement noise, while the plateau at the large lag values is proportional to the variance in the signal fluctuations at time lags longer than the variability timescale of the underlying physical process. The structure function is usually characterised by the slope of the power law portion, given by $d~\log SF/d~\log\tau$, which is a measure of the variability of the process.

The Fig.~1 shows the structure function plots for the observations not affected by atmospheric events. The rapid drop in the values of SF at very large time lags can be explained by the limited sampling of the data at large time lags. The irregular bends in the curves can be attributed to irregular and uneven sampling of data. The structure function curves suggest that most of the power lies in the longer timescale fluctuations ($\sim 60 - 130 minutes$) in the Sgr A* flux. The slopes of the power law are calculated for the long timescale lags. The values of the slope are found to be $\sim$ $0.4 - 0.7$ which can be described by pink noise. On three occasions, we detect the slope $\sim 1.0$ suggesting strong variability at those lag time scales, an indication of flare events.

\begin{figure*}
\centering
\resizebox{0.95\textwidth}{!}{
\subcaptionbox{\Huge}{\includegraphics[width=0.70\textwidth]{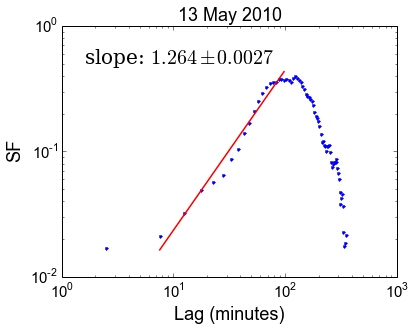}}
\subcaptionbox{\Huge}{\includegraphics[width=0.70\linewidth]{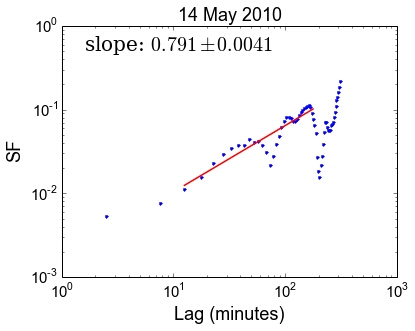}}
\subcaptionbox{\Huge}{\includegraphics[width=0.70\linewidth]{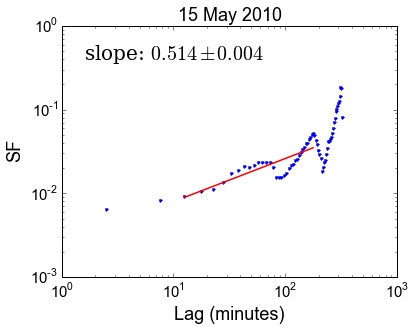}}}
\vspace{1pt}
\resizebox{0.95\textwidth}{!}{
\subcaptionbox{}{\includegraphics[width=0.70\textwidth]{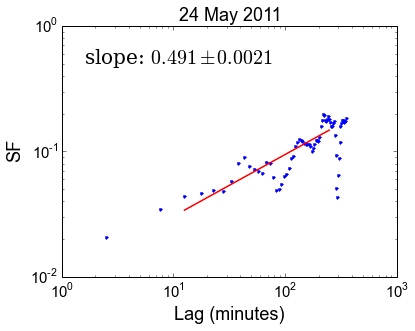}}
\subcaptionbox{}{\includegraphics[width=0.70\linewidth]{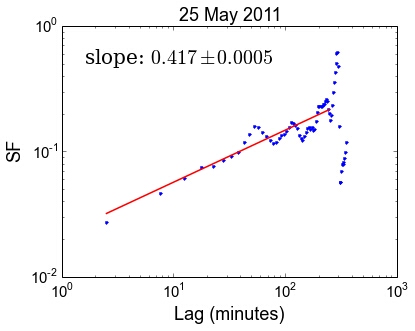}}
\subcaptionbox{}{\includegraphics[width=0.70\linewidth]{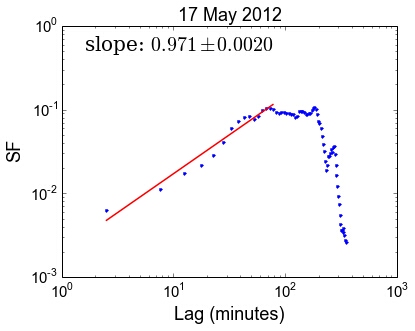}}}
\vspace{1pt}
\resizebox{0.95\textwidth}{!}{
\subcaptionbox{}{\includegraphics[width=0.70\textwidth]{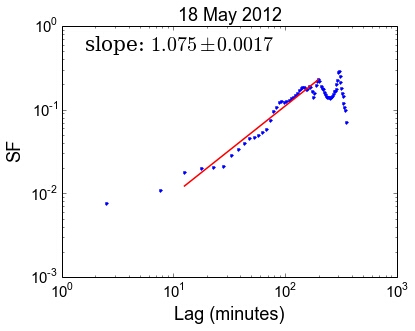}}
\subcaptionbox{}{\includegraphics[width=0.70\linewidth]{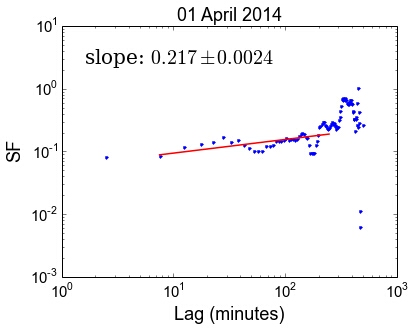}}
\subcaptionbox{}{\includegraphics[width=0.70\linewidth]{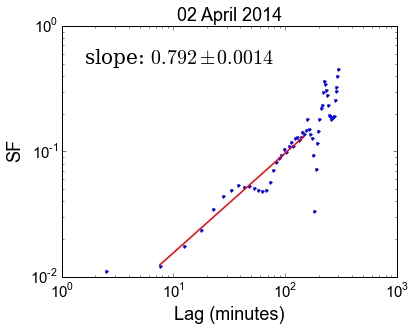}}}
\vspace{1pt}
\resizebox{0.60\textwidth}{!}{
\subcaptionbox{\tiny}{\includegraphics[width=0.30\textwidth]{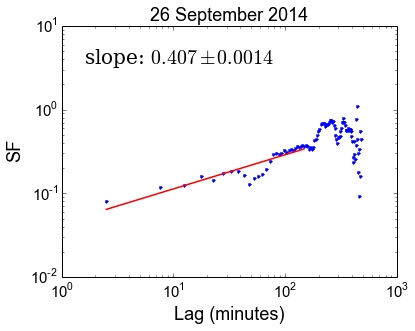}}
\subcaptionbox{\tiny}{\includegraphics[width=0.30\linewidth]{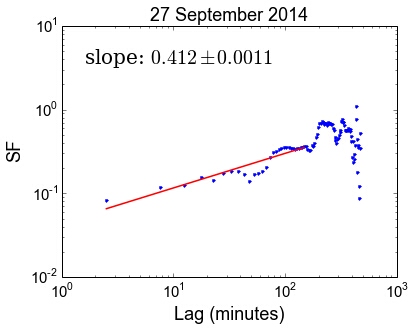}}}

\label{fig_SF}
\caption{The structure function plots for the observations. Y-axis shows the SF in the units of $Jy^2$.}
\end{figure*}


\subsection{Bayesian Blocks Analysis}

To evaluate the light curve of Sgr A* and to search for flare events, we make use of the Bayesian Blocks algorith. This algorithm assumes that the light curve can be divided into series of blocks of constant flux, with each block having flux values with no significant variation. Larger variations in the flux will form separate blocks with different flux values. The points separating each block are called change points. We make use of the ``point measurements'' Bayesian Blocks algorithm described by \citealt{Scargle13}, which is used for the analysis of radio data. The number of blocks and change points are determined using the false positive probability $p0$ ($=0.05$), and then iteratively determining the optimum number of blocks. This method has been used extensively to analyse the observations of X-ray and Gamma-ray sources \citep{Ackermann13,Burgess14,Rama14}, and X-ray observations of Sgr A* \citep{Neilsen13, Barri14, Ponti15, Mossoux15}.

The flares are defined by the Bayesian blocks with flux density values significantly different than the quiescent values. The start and end times of the flares are given by the change points describing the blocks with elevated flux density values, with the flare duration given by the total duration of all the blocks. The peak of the flare is calculated from the average flux density value in the block where peak is observed. We also take into account the characteristic timescale of $\sim 100-150$ minutes of the flare events at 3mm wavelength \citep{Mauerhan05} to determine possible flare events.

We observed flaring events on three occasions, 13 May 2010, 17 May and 18 May in 2012. The light curves of the corresponding dates are shown in Fig. 2. These flares will be discussed in the following section in detail.

\begin{figure*}
\centering
\resizebox{0.95\textwidth}{!}{
\subcaptionbox{\large{13 May 2010}}{\includegraphics[width=0.70\textwidth]{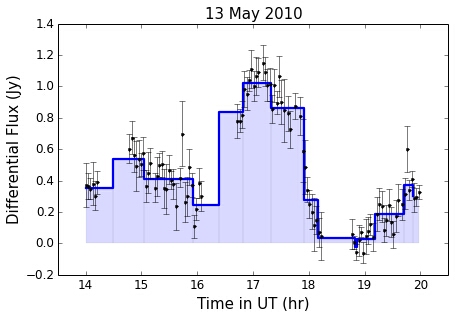}}
\subcaptionbox{\large{17 May 2012}}{\includegraphics[width=0.70\linewidth]{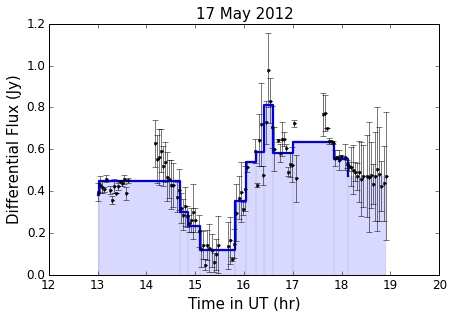}}}
\vspace{1pt}
\resizebox{0.95\textwidth}{!}{
\subcaptionbox{\large{18 May 2012}}{\includegraphics[width=0.70\textwidth]{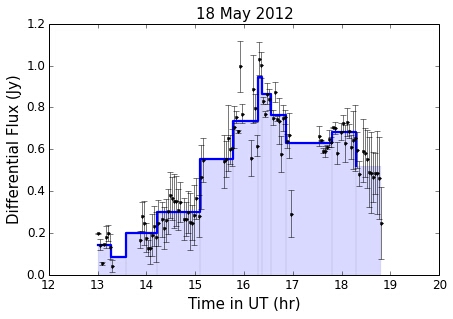}}
\subcaptionbox{\large{25 May 2011}}{\includegraphics[width=0.70\linewidth]{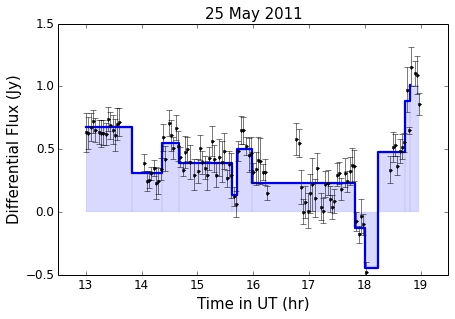}}}
\label{fig1}
\caption{The differential light curves of Sgr A* with ATCA at 3mm for the days during which flares were detected (black points), along with the Bayesian blocks representation of the light curves (blue shaded region). A differential light curve from 25 May 2011 observation is shown as an example of light curve with no strong flare activity. No flares were observed during that day with their start and stop times within the observed time range.}
\end{figure*}

\subsection{Flare events}

\noindent\textbf{13 May 2010:}
\vspace{0.2cm}

\noindent The first two flare events were detected on 13 May 2010. The first flare was observed at the beginnig of observations, starting at around 13.9 hrs and ending at 15.8 hrs UT.\@ The differential flux of Sgr A* increased by 0.5 Jy in 50 minutes, before decreasing slowly to 0.1 Jy. The second flare was observed immediately after the first, where the flux increased again steadily from 0.2 Jy to 1.0 Jy in 1 hr, followed by a decline back to 0 Jy. This was one of the strongest flare detected from all the observations with a flux variation of 0.8 Jy.

\vspace{0.2cm}

\noindent\textbf{17 May 2012:}
\vspace{0.2cm}

\noindent We detected two instances of strong variation in differential flux on 17th May, 2012, starting at the beginnig of observations and at 15.6 hrs UT respectively. The differential flux first increased by 0.4 Jy within 1 hour, followed by a decrease to 0.1 Jy. The second flare followed with a flux increase to 0.8 Jy in 45 minutes. The two flares are centred at 14.0 hr UT and 16.5 hr UT respectively.
\vspace{0.2cm}

\noindent\textbf{18 May 2012:}
\vspace{0.2cm}

\noindent A strong flare was detected on 18 May 2012 starting at 15.0 hr UT, where the differential flux increased by 0.7 Jy within 1.5 hrs before subsiding by 0.4 Jy. A second, slightly weaker flare was observed, centred at 18.0 hr UT, overlapping the declining part of the first flare, where there was a differential flux increase by 0.3 Jy in 45 min before decreasing to 0.0 Jy.

\vspace{0.2cm}
\noindent\textbf{The 2013 and 2014 data:}
\vspace{0.2cm}

The observations for 2013 and 2014 were scheduled to monitor the activity of Sgr A* during the periapse passage of the dusty S-cluster object (DSO) which was discovered recently, where large flux variations were predicted (\citealt{Eckart13, Valencia15}, see Section~\ref{subsection:DSO} for details). During the 2013 campaign, observations were not possible for 3 of the 6 scheduled days due to heavy rainfall. On the rest of the three days, parts of the observation were affected by atmospheric turbulence, but we detected no significant excursion in the unaffected parts of the light curves. In 2014, the observations went without any hurdles and the signal to noise ratio was good. We do not detect any significant flux variation on the flare time scales of $\sim 150$ minutes and no flares were observed with their start and end times within the observed time range.

\vspace{0.2cm}

During all the observed flaring events, the calibrator flux density has remained constant, thus supporting our conclusion that the observed flux density variations originate from Sgr A*.

\section[]{Flare analysis}

To explain the flaring activity of Sgr A*, several models have been proposed. Here, we use the adiabatically expanding plasmon model to analyse the flares.

\subsection{The Plasmon Model}
\label{subsection:Adiabatic}

The expanding plasmon model proposed by \citet*{Laan66} has been used to explain the radio and sub-mm flares of Sgr A*.
Previous studies of multiwavelength analysis of the simultaneous flaring of Sgr A* \citep{Eckart06a, Eckart08b, Eckart12, YZ06a, YZ08} have shown that the light curves peak at successively lower frequencies (from NIR to sub-mm to millimeter to radio frequencies) as self absorbed synchrotron source expands after the initial event that energises the electrons. The peak of the emission at a particular frequency occurs when the blob becomes optically thin at that frequency. The size of the blob determines the peak of the flux while the expansion speed determines the time period of the flare. The model predicts that at high frequencies where the blob is optically thin, the simultaneous flaring and declining emission occurs at NIR and X-ray wavelengths, and then successively at lower frequencies in sum-mm and mm radio regime where the blob is initially optically thick.

This model assumes that a uniformly expanding blob of relativistic electrons with a power-law energy spectrum, $n(E) \propto E^{-p}$ embedded in a magnetic field that declines as $R^{-2}$. The particle energy of each relativistic particle declines as $R^{-1}$ and the particle density scales as $R^{-3}$, as a result of expansion of the blob \citep*{Laan66}. 
The frequency dependence of the flux density $I_{\nu}$ is given as
\vspace{0.1cm}

\begin{equation}
    {I_\nu}  =  I_0 {\tilde{\nu}}^{5/2} \rho^3 \xi({\tau}_{\nu})
    \label{eq:Snu}
\end{equation}

\vspace{0.1cm}
Here $\tilde{\nu} = (\nu/\nu_0)$ is the normalised frequency, $\rho = ({R}/{R_0})$ is the expansion factor for the expanding blob of synchrotron emitting relativistic electrons with $R_0$ as the initial radius, and $\xi({\tau}_{\nu}) = \frac{1 - \exp({-\tau_{\nu}})}{1 - \exp({-\tau_0})}$, which describes the exponentially declining part of the flare. The optical depth of the synchrotron radiation  $\tau_{\nu}$ scales as
\vspace{0.1cm}
\begin{equation}
    \tau_\nu = {\tau_0} {\tilde{\nu}}^{-(p+4)/2} \rho^{-(2p+3)}
    \label{eq:taunu}
\end{equation}

\vspace{0.1cm}
where $\tau$$_0$ is the critical optical depth at the peak of the flux density at any particular frequency, and $\nu_0$ is the frequency at which the flux density peaks when $R = R_0$ at time $t = t_0$. For frequencies $\geq$ $\nu_0$ the blob is initially optically thin. The initial optical depth $\tau$$_0$ depends only on the particle spectral index $p$ through the condition
\vspace{0.1cm}

 \begin{equation}
    e^{\tau_0} - (2p/3 + 1)\tau_0 - 1 = 0.
    \label{eq:tau0}
\end{equation}
\vspace{0.1cm}
 
Thus for a given particle spectral index $p$ and peak flux density $I_0$ at the turnover frequency $\nu$$_0$, the model gives the variation in flux density at any frequency as a function of the expansion factor $\rho~=~(R/R_0)$. To fit the model to the light curves, we have to convert the dependence of $I_{\nu}$ and $\tau_{\nu}$ from expansion factor to time. We assume that the blob expands uniformly at a constant speed so that the expansion radius $R$ is a simple linear function of time, with a constant expansion speed of $v_{\mathrm{exp}}$, given by $R-R_0~=~v_{\mathrm{exp}} (t-t_0$).  The initial radius $R_0$ is usually expressed in multiples of the Schwarzschild radius. One Schwarzschild radius corresponds to $R_{\mathrm{s}}$ = 2$GM$/$c$$^2$ $\sim$ 10$^{10}$$\,$m for a supermassive black hole of mass $\sim~4 \times~10^6 M_{\odot}$, which translates to the velocity of light to be about 100 $R_{\mathrm{s}}$ per hour. At times $t$$<$$t_0$, we assume that the source has an optical depth equal to its frequency-dependent initial value $\tau$$_0$ at $R$=$R_0$. For the optically thin part of the spectrum, the flux increases initially with increasing source size at constant optical depth $\tau$$_0$ and then decreases with decreasing optical depth as it expands. For $t$$>$$t_0$, an increase in the turnover frequency $\nu$$_0$ or the initial source size $R_0$ shifts the decaying flank of the curve to later times, thus lengthening the duration of the flare. A decrease in the spectral index $\alpha_{\mathrm{synch}} = (p-1)/2$ or peak flux density $I_0$ also has the same effect in the shape of the decaying flank of the curve, while increasing the adiabatic expansion velocity, $v_{\mathrm{exp}}$, shifts the peak of the light curve to earlier times. Flare timescales are longer at lower frequencies and have a slower decay rate, as a result of adiabatic expansion. 
\vspace{0.2cm}

For the modelling of the NIR flares, the synchrotron self-Compton (SSC) model has been used in conjugation with the adiabatically expanding plasmon model. In this model, the synchrotron and SSC components in the temporary accretion discs are upscattered by SSC processes to NIR/X-ray regime and then the adiabatic expansion results in the radio/sub-mm emission. The detailed formulation of this model has been discussed by \citet{Sabha10} and \citet{Eckart12}. This formalism allows us to constrain the strength of the magnetic field  as a function of the turnover frequency $\nu_m$, which is given by $B \sim \theta^4 \nu_m^5 I_m^{-2}$. Here $\theta$ is the angular size of the synchrotron component, while $I_m$ is the peak flux density at the turnover frequency.

\begin{table*}
  \centering
  \begin{minipage}{140mm}
   \caption{The parameters for the individual source components estimated from the adiabatically expanding plasmon model.}\label{table:Table2}
    \begin{tabular}{c c c r r r r r r}
      \hline
      Date & Flare & $v_{\mathrm{exp}}$ & $S_{\mathrm{max}}$ & $\nu_{\mathrm{max}}$ & $R_0$ & $\alpha_{\mathrm{synch}}$ & $\chi^2$ \\
      & & [$c$] & [Jy] & [GHz] & [$R_s$] & & [G] & \\
      \hline
      \hline
      13 May 2010 & A & $0.015 \pm\ 0.002$ & $3.57 \pm\ 0.21$& $469.2 \pm\ 70.0$ & $1.67 \pm\ 0.06$ & $0.60 \pm\ 0.15$ &  1.43 \\
      & B & $0.021 \pm\ 0.001$ & $6.95 \pm\ 0.29$ & $378.73 \pm\ 102.5$  & $3.61 \pm\ 0.03$ & $0.77 \pm\ 0.20$ & 1.94 \\
      & & & & & & & & \\
      17 May 2012 & C & $0.0135 \pm\ 0.002$ & $5.19 \pm\ 0.37$ & $542.7 \pm\ 67.4$ & $1.60 \pm\ 0.05$ & $0.66 \pm\ 0.21$ & 1.76 \\
      & D & $0.022 \pm\ 0.003$ & $7.45 \pm\ 0.27$ & $512.3 \pm\ 57.1$ & $1.67 \pm\ 0.03$ & $0.675 \pm\ 0.30$ & 2.06 \\
      & & & & & & & & \\
      18 May 2012 & E & $0.023 \pm\ 0.004$ & $7.73 \pm\ 0.4$  & $545.9 \pm\ 72.0$ & $2.27 \pm\ 0.01$ & $0.55 \pm\ 0.32$ & 1.98 \\
      & F & $0.017 \pm\ 0.003$ & $6.25 \pm\ 0.33$ & $652.0 \pm\ 81.5$ & $1.35 \pm\ 0.02$ & $0.67 \pm\ 0.23$ & 2.16 \\
      \hline
    \end{tabular}

  \end{minipage}
\end{table*}

\vspace{0.2cm}

\noindent\textbf{The modelling of the 13 May 2010 flares:}
\vspace{0.2cm}

\noindent The first flare (flare A) is modelled using a source component of initial radius $1.67$ $R_{\mathrm{s}}$. The expansion speed of the blob was found to be $0.015$ $c$. The particle spectral index $p$ and the turnover frequency were 2.2 and 469.2 GHz respectively.

The second flare (flare B) is modelled using a component arising at 15.8 hr UT and centred at 17.2 hr UT.\@ The fitted model gives the expansion speed of $0.021~c$ for a source of initial radius of $3.61~R_{\mathrm{s}}$ and particle index of $2.54$. The turnover occurs at $378.7$ GHz.

\vspace{0.2cm}

\noindent\textbf{The modelling of the 17 May 2012 flares:}
\vspace{0.2cm}

\noindent We detected two flaring events on this day. The first source component (flare C)  originate at 12.8 hr UT and reaches the peak intensity at 14.15 hr UT.\@ The expansion speed of the blob of initial radius 1.60 $R_{\mathrm{s}}$ was estimated to be 0.0135 $c$ with the particle spectral index of 2.32 and the turnover frequency of 542.7 GHz.

For the second flare (flare D), the source component is assumed to have originated at 15.6 hr UT, and reaching peak flux density at 16.5 hr UT.\@ The expansion speed for the blob is 0.022 $c$, with the particle index of 2.35. The turnover frequency was found to be 512.3 GHz for initial radius of $1.67$ $R_{\mathrm{s}}$.
\vspace{0.2cm}

\noindent\textbf{The modelling of the 18 May 2012 flare:}

\noindent The first of the two detected flare events was modelled with the source component of radius 2.27 $R_{\mathrm{s}}$ with the expansion speed of $0.023$ $c$, particle index of 2.1 and the turnover frequency of 545.9 GHz. The modelling of the second flare required a source component of size $1.35 R_{\mathrm{s}}$ expanding at the speed of $0.017$ $c$. The spectral index and the turnover frequency of the blob was estimated to be 2.34 and 652.0 GHz respectively.

\begin{figure*}
\centering

\resizebox{1.0\textwidth}{!}{
\subcaptionbox{\LARGE{Flare A $-$ 13 May 2010}}{\includegraphics[width=0.85\linewidth]{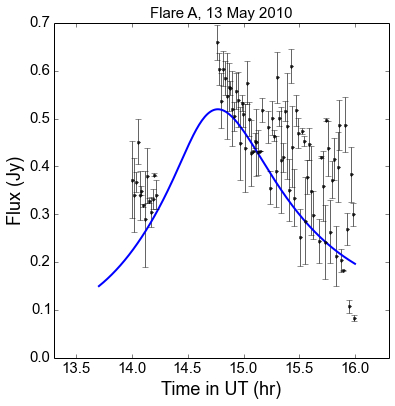}}
\subcaptionbox{\LARGE{Flare B $-$ 13 May 2010}}{\includegraphics[width=0.89\linewidth]{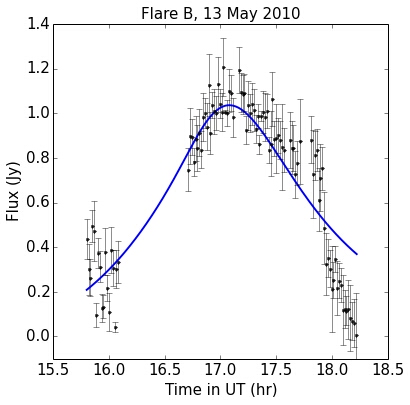}}
\subcaptionbox{\LARGE{Flare C $-$ 17 May 2012}}{\includegraphics[width=0.85\linewidth]{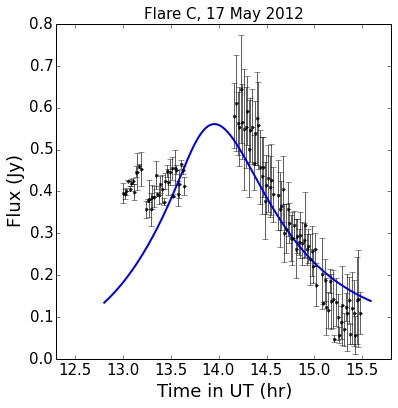}}
}

\resizebox{0.85\textwidth}{!}{

\subcaptionbox{{Flare D $-$ 17 May 2012}}{\includegraphics[width=0.40\linewidth]{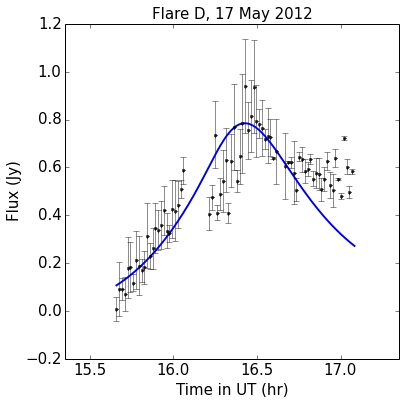}}
\subcaptionbox{{Flare E $-$ 17 May 2012}}{\includegraphics[width=0.50\linewidth]{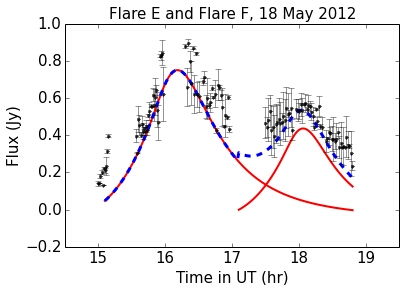}}}


\caption{The fitted models for the detected flares. The black data points denote observed differential flux points. The individual flares are shown by blue curve. In case of overlapping flares, the individual flares are shown by red curve and the blue dotted curve shows the combined fitted model.}
\label{ModelFit}
\end{figure*}

The details of all the fitted parameters from the expanding plasmon model for all the observed flare events can be found in Table~\ref{table:Table2}. The modelled curves fitted to the differential light curve data are shown in the Fig.~\ref{ModelFit}. 

\vspace{0.2cm}

\subsection[]{Modelling Context}

The adiabatically expanding plasmon model has been used extensively for the analysis of flaring of Sgr A*. It successfully describes the evolution of the synchrotron-SSC blob that gives rise to the flaring event and its physical parameters. But this model does not explain the origin of the expanding plasmon. Two models, viz.\@ the hot-spot model and the jet model, have been invoked to explain the origin of the synchrotron-SSC blob. These models can be briefly described as follows.

\noindent\textbf{The Hot Spot model:}
\vspace{0.2cm}

\noindent A frequently used model to explain the short term NIR and X-ray variability is the orbiting hot spot model (\citealt{Brod-Loeb06, Eckart06a}). This model consists of bright overdensities forming hot spots in the accretion disk that are revolving around the SMBH in a Keplerian orbit close to the last stable orbit. The relativistic beaming effects and Doppler shift causes the component to look brighter while approaching than while receding. The relativistic effects, combined with the SSC modelling are able to estimate the NIR and X-ray flare parameters well \citep{Eckart06a, Zamani10}. The hot spot model has been applied to the radio and sub-mm flares with the radiatively inefficient accretion models by accounting for the disc opacity effects in the radio band (\citealt{Brod-Loeb06, Li09}). The hot spot model can explain some aspects of the accretion discs in active galactic nuclei and accreting stellar-mass black holes in binary systems, but a standard gaseous disc is absent near Sgr A* (\citealt{Levin}). Though this model can explain the flaring activity at NIR/X-ray and radio/sub-mm wavelengths, it does not account for the $\sim 100$ minute time lag between flares in the NIR and radio regime.

The time variable NIR emission from Sgr A* can be described as a result of a single continuous power-law process \citep{Witzel12}. This continuous variability process shows a break time scale around a few hundred minutes. It is also on these time scales that the Fourier power in the variability amplitudes drops. Therefore extreme flux density excursions typically last for about 100 minutes, often considered as flares. There is an indication that a similar time scale may be of importance in the radio domain. \citet{Mauerhan05} find an indication of a 150 minute time scale for Sgr A* at 3 mm wavelength.
\vspace{0.2cm}

\noindent\textbf{The Jet Model:}
\vspace{0.2cm}

\noindent Another model that has been used to study the flare emission of Sgr A* is the Jet-ADAF model (\citealt{Falcke93, Markoff01, Yuan02}). The combined accretion-jet model has been used to describe the broadband spectrum and the quiescent emission of Sgr A*. The interaction of the synchrotron and SSC components in the accretion disc with the short jet gives rise to the emission from Sgr A*. To describe the flaring this model assumes a source component of synchrotron emitting relativistic electrons in the accretion flow that gets caught up in the nozzle/outflow of the jet. The upscattering of the SSC components give rise to X-ray and NIR flares. The blob gets accelerated and expands as it moves along the jet. Thus different sections of the jet give rise to emission at successive radio wavelengths. The Jet-ADAF model successfully predicts the simultaneous flaring in the NIR and X-ray regime, and explain the time delay between the NIR/X-ray and radio/sub-mm flares. It also predicts the 10\% linear polarisation as observed in the sub-mm bump. The jet model has been successfully applied to several low luminosity AGNs (LLAGN) (\citealt{Herrn98, Bietenholz00, Yuan03ANS, Anderson04, Liu13, Nemmen14}). Sgr A*, considered to be the weakest LLAGN, is also assumed to have a jet outflow. Although a jet from Sgr A* has been predicted, there have not been robust detections of any jet from Sgr A*. There has been circumstantial evidence for the presence of a jet. Detections of the extended features at position angles 120$^o$ to 130$^o$ (\citealt{Eckart06a, Eckart06b, Morris04}) coupled with the NIR polarization of synchrotron radiation suggests the presence of a jet (\citealt{Shahzamanian15}). Comparative analysis of the AGN parameters gives an upper limit on the size of the Sgr A* jet to be around $10^{-3}$ AU, which is significantly smaller than the current best resolution (\citealt{Markoff03}). It is possible that the low surface brightness of the jet makes its detection difficult and it remains hidden. Future observations with higher angular resolution with the next generation telescopes will be able to resolve the close vicinity of Sgr A*.

\vspace{0.2cm}

It is possible that the two qualitatively disparate approaches will be united (\citealt{Eckart15}). In the simplest approach, the relativistic bulk motion of the orbiting or the outward travelling source is described by the Lorentz factor. A helical-type trajectory of the source is in fact more likely. While in the central plane the flux variations are assumed to be the result of the orbital motion, in the case of an outflow the variations are related to ejection directed along the jet (with bulk velocity close to the speed of light). This is because the foot-points of magnetic field lines are embedded into the orbiting material near the black hole, where they are frozen in the accreted medium and dragged in the toroidal direction, while further out the emerging field-lines are twisted into the spiral shape that induces the radial component of the blob velocity.

A phenomenological description of the orbiting blobs in helical motion along a funnel or cone shaped magnetic surfaces anchored in the inner accretion disc was recently reexamined by \citet{Mohan15}. It has been proposed that quasi-periodic variability can be caused by motion along outward spiralling trajectories. In principle, this can be tested by observing the variable signal in different energy bands, and by polarimetry \citep*{Bro09}.

\section[]{Discussion}

The adiabatically expanding plasmon model has been successfully used to explain the flaring of Sgr A* at the NIR, submillimeter and radio wavelengths. \citep{Eckart06a, Eckart08b, YZ06a, Kunneriath10}. These observations show that the synchrotron emission is initially optically thick at millimeter wavelengths, and optically thin at NIR wavelengths, with the turnover occurring in the submillimeter region. The expansion modelling of the millimeter flares by \citet{YZ06b, YZ08} predicts the turnover occurring at frequencies $\geq 350$ GHz. Similar values of the turnover frequency of around few hundred GHz are estimated by \citet{Kunneriath10} and \citet{Eckart12}. The values of the initial blob radius $R_0$ and the particle spectral index $p$ strongly depend on the value of the turnover frequency, which determine the flux density in the optically thick and optically thin part of the spectrum, respectively. Higher values of $\nu_0$ in the THz regime also give reasonable fits to the data but they overestimate the particle spectral index and the peak flux densities at submillimeter wavelengths \citep{Eckart08b}. The particle index of 1.5 $\leq$ $p$ $\leq$ 3.5 best describes the flux density at millimeter wavelength \citep{YZ06a}. The estimates of the parameters obtained from our data are fairly consistent with these values.

We obtain the expansion velocities of the blob  of $0.013 - 0.025$ $c$ which are within the range of previously published results \citep{Eckart08b, YZ08, Li09}. These values are an order of magnitude smaller than the relativistic sound speed in the absence of external pressure in the vicinity of the SMBH, which is estimated to be around $0.2$ $c$. Such low expansion speeds imply that the expanding blob either has a higher bulk motion or it is confined to the vicinity of Sgr A* in the form of a disc or a corona. An expansion of a hot spot source component through shearing caused by the differential rotation within the accretion disc may explain the low expansion velocities. We also find that the magnetic field strength values obtained using the SSC modelling constraints are generally less than 10 $G$, which are within the range of previously observed values \citep{Kunneriath10, Li09}.

\begin{figure}
\centering
\includegraphics[width=0.45\textwidth]{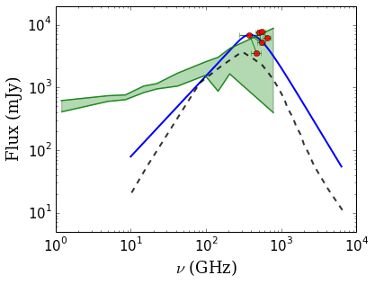}
\caption{The SED of Sgr A*, with the observed flux density values from literature shown by green shaded region, the modelled SED by \citet{Dexter10} shown by black dashed line. The red circles represent the peak flux density at the turnover frequency obtained for the observed flares. The solid blue line shows the spectrum of one of the modelled flares (flare B).}
\label{SED}
\end{figure}

Fig. 4 shows the plot of spectral energy distribution (SED) of Sgr A*. The green shaded region describes the extent of the observed flux density values of Sgr A* from literature \citep{Zylka95, Serabyn97, Pierce-Price00, Aitken00, Marrone03}, while the black dashed line shows the model fit for the SED given by \citet{Dexter10}, which gives a bound to the observed flux density values. The red points represent the peak flux density $S_{\mathrm{max}}$ obtained at the turnover frequency $\nu_0$ from the modelling of our observations. We show the spectrum for flare B (blue solid line) as an example. As it can be seen from the peak flux values that we obtain are consistent with the other observations within the margin of error. The spectrum of the modelled flare has a distribution compatible with previous observations and with the SED model by \citet{Dexter10}. The spectral index obtained for our modelled flares is consistent with the observed flares in radio \& sub-mm wavelengths \citep{YZ06a, YZ08, Li09, Kunneriath10} and with the observed infrared spectral indices and the associated infrared flux is well within the range of observed values.

\subsection[]{No detection of DSO-induced activity}
\label{subsection:DSO}

Recently, a fast moving infrared excess source was detected approaching the central black hole, called G2/DSO (dusty S-cluster object) \citep{Gillessen12, Gillessen13a, Eckart13}. This has raised attention with regards to the GC and sparked many multiwavelength observation campaigns of the GC.\@ Several theories have been put forward to explain the origin and nature of the object. From the infrared excess of the source, it has been speculated that it consists of dominant fraction of gas and dust \citep{Gillessen12, Gillessen13a, Pfuhl15}, while the NIR/MIR spectrum allow for the stellar contribution which has been favoured by others (\citealt{Eckart13, Murray-Loeb12, Scoville13, Ballone13, Phifer13, Jalali14}). From the spectroscopic measurements, the object was predicted to undergo peribothron around in 2014. \citet{Pfuhl15} detect a blue-shifted Br$\gamma$ emission from the DSO in their April 2014 dataset, while \citet{Valencia15} detect blue-shifted emission in their dataset after May 2014, where they conclude that the DSO has remained spatially compact. The L' band observations by \citet{Witzel14} also find that the DSO has maintained the compact shape and its Keplerian orbit.

The flyby of the DSO is predicted to affect the flaring activity of Sgr A* due to the extra amount of cold material reaching the SMBH vicinity. Computational models suggest that if the DSO develops a bow shock then it may lead to shock acceleration of electrons and correspondingly strong flares in radio emission. The size of the bow shock determines the strength of the flares. The initial estimates predicted that the excursions in the flux at centimeter wavelengths would be of the order of $1-20$ Jy. The revised estimates put the values of the order of $0.01-0.2$ Jy (\citealt{Narayan12, Sadowski13, Crumley13}), which are within the range of the normal flux variations of Sgr A*.

So far, several observational campaigns at different wavelengths did not reveal any deviation from the usual flaring activity of Sgr A*. The NIR observations by \citet{Valencia15} with SINFONI during $2013-2014$ observational campaigns detect three flares of few mJy strength, with no exceptional flaring activity. No increased continuum flux density or extraordinary variability detected in the $\geq$ 2 keV band in X-ray regime \citep{Haggard14}, which would be expected from a shock-heated gas. Monitoring of the GC with VLA has not resulted in detectable strong radio flares, and the current estimated variability would be within the normal range of flux density variations observed towards Sgr A* (\citealt{Sjou14}).

Our observations of the Galactic Centre with ATCA in $2013-2014$ do not detect any significant variation in the flaring activity of Sgr A*. The observations taken during the approach of the DSO towards peribothron did not detect any flux density variation in the light-curves within the characteristic flare time-scales. This is consistent with the compact blue-shifted Br$\gamma$ emission of the DSO observed in the NIR imaging. A presence of a higher mass stellar central source embedded in the DSO would imply that the object would retain compactness through the flyby and the material surrounding the core would be largely unaffected and would not cause large excursions in flux density of Sgr A*. On the other hand, if the central source has the mass of about one solar mass, it may allow a large part of the gas and dust to fall onto Sgr A*. However, for a very short time scale right after the peribothron passage, it is uncertain whether the gas close to the star will remain in the Roche lobe of the star. Given the short distance of approach of the DSO during the peribothron around the SMBH, it is unclear whether the flyby would lead to any flaring activity.

\section[]{Summary}

We have presented the results of the 3mm observations of the GC carried out with ATCA between 2010 and 2014. The light curves of Sgr A* were obtained by calculating the differential flux density using the median subtracted flux density of two orthogonal pairs of baselines. From this data, we detect six different flaring events over three observation days. To model these flares we have used the expanding plasmon model which involves adiabatic expansion of synchrotron source components, based on a model by \citet*{Laan66}. The model parameters values, such as the initial blob radius, expansion velocities, spectral index and such were obtained from the fitting. These values are consistent with the previous studies. We estimate the expansion velocities within the range of 0.014 $c$ $-$ 0.025 $c$. The turnover frequency $\nu_0$, separating the optically thick millimeter and sub-mm region ($\nu < \nu_0$) and the optically thin near-IR and X-ray region ($\nu > \nu_0$) was found to lie in the sub-mm region between 350 GHz and 660 GHz. The light curve structure is consistent with the previously found variability of Sgr A*, which can be interpreted as emission from the material in relativistic orbits around the SMBH.\@

Our observations in 2013 and 2014 were taken during the expected peribothron passage of the dusty infrared excess source, known as DSO/G2. The flyby of the object was predicted to give rise to enhanced flaring activity of Sgr A*. During our observations, we do not detect any significant flaring activity, as expected from the observed compact nature of DSO after the peribothron. Further observations of the GC are required to monitor the probable induced flaring activity due to the flyby of the DSO.\@

\section*{Acknowledgements}

We would like to thank the anonymous referee for their constructive and helpful comments. 

This work was supported in part by the Deutsche Forschungsgemeinschaft (DFG) via the Cologne Bonn Graduate School (BCGS), the Max Planck Society through the International Max Planck Research School (IMPRS) for Astronomy and Astrophysics. A. Borkar, B. Shahzamanian, and N. Sabha are members of the IMPRS.\@ Part of this work was supported by fruitful discussions with members of the European Union funded COST Action MP0905: Black Holes in a Violent Universe and the Czech Science Foundation DFG collaboration (No. $13-00070J$) and with members of the European Union Seventh Framework Program (FP7/$2007-2013$) under grant agreement no 312789; Strong gravity: Probing Strong Gravity by Black Holes Across the Range of Masses.

\bibliographystyle{mn2e}
\bibliography{draft0}

\appendix

\section[]{3 mm differential flux density curves of Sgr A*}

The individual differential flux density curves at 3mm for all the observations from 2010 to 2014 are provided here.

\begin{figure*}
\centering

  \begin{multicols}{2}
    \includegraphics[width=0.9\linewidth]{./A_Borkar_ATCA_accepted_Images/BB_20100513_.jpg}\par\caption{13 May 2010}
    \includegraphics[width=0.9\linewidth]{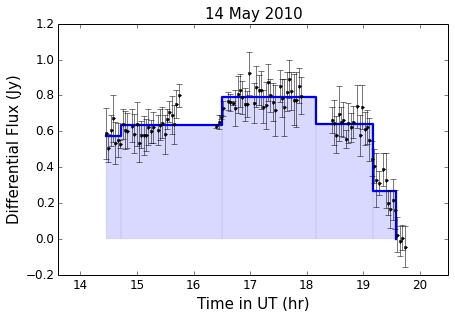}\par\caption{14 May 2010}
  \end{multicols}
  \begin{multicols}{2}
    \includegraphics[width=0.9\linewidth]{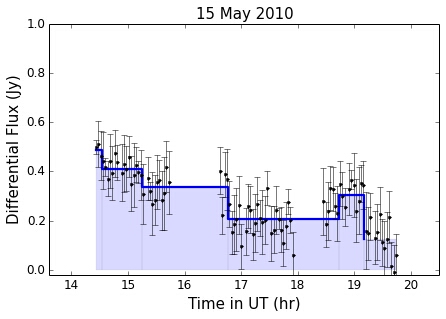}\par\caption{15 May 2010}
    \includegraphics[width=0.9\linewidth]{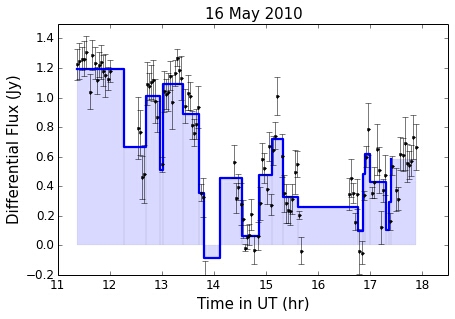}\par\caption{16 May 2010}
  \end{multicols}
  \begin{multicols}{2}
    \includegraphics[width=0.9\linewidth]{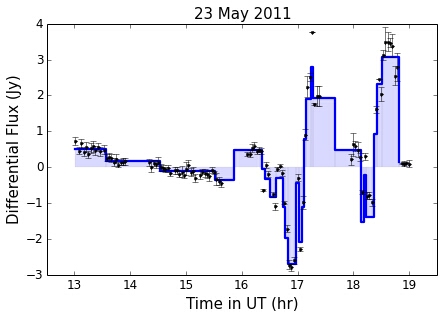}\par\caption{23 May 2011}
    \includegraphics[width=0.9\linewidth]{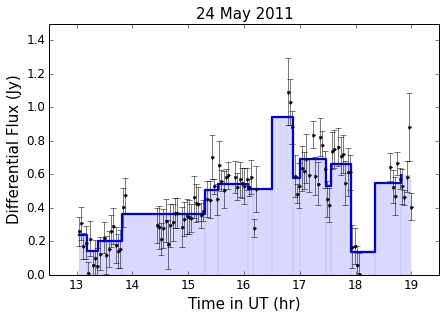}\par\caption{24 May 2011}
  \end{multicols}
  \begin{multicols}{2}
    \includegraphics[width=0.9\linewidth]{./A_Borkar_ATCA_accepted_Images/BB_20110525_.jpg}\par\caption{25 May 2011}
    \includegraphics[width=0.9\linewidth]{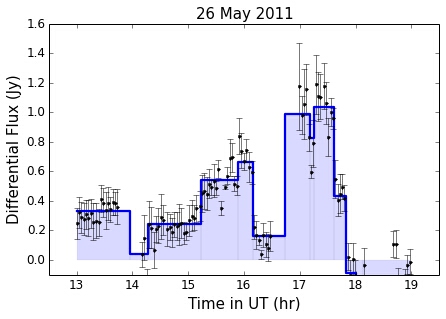}\par\caption{26 May 2011}
  \end{multicols}
\end{figure*}

\begin{figure*}
  \begin{multicols}{2}
    \includegraphics[width=0.9\linewidth]{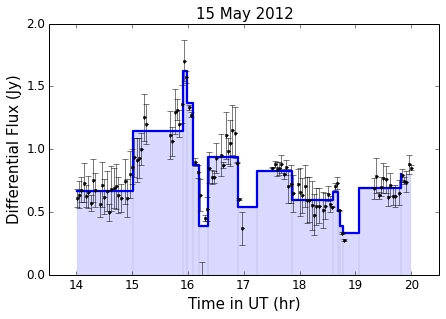}\par\caption{15 May 2012}
    \includegraphics[width=0.9\linewidth]{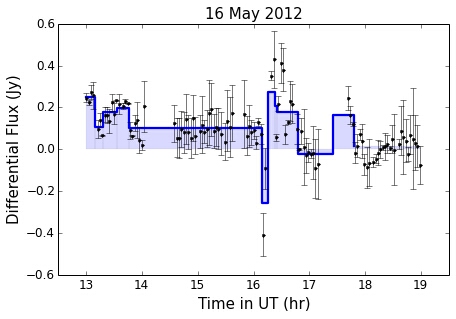}\par\caption{16 May 2012}
  \end{multicols}
  \begin{multicols}{2}
    \includegraphics[width=0.9\linewidth]{./A_Borkar_ATCA_accepted_Images/BB_20120517_.jpg}\par\caption{17 May 2012}
    \includegraphics[width=0.9\linewidth]{./A_Borkar_ATCA_accepted_Images/BB_20120518_.jpg}\par\caption{18 May 2012}
  \end{multicols}
  \begin{multicols}{2}
    \includegraphics[width=0.9\linewidth]{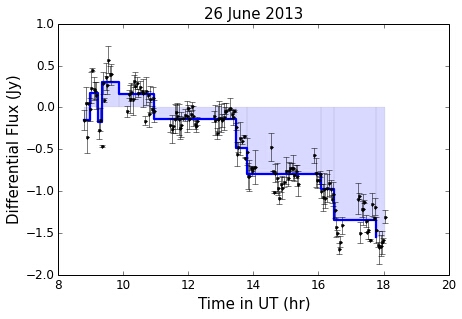}\par\caption{26 June 2013}
    \includegraphics[width=0.9\linewidth]{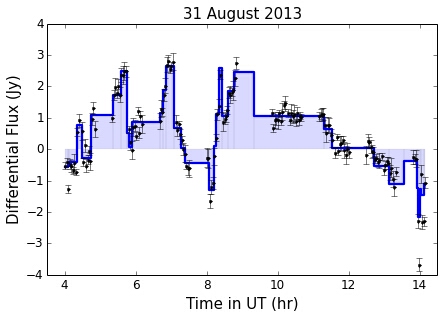}\par\caption{31 August 2013}
  \end{multicols}
  \begin{multicols}{2}
    \includegraphics[width=0.9\linewidth]{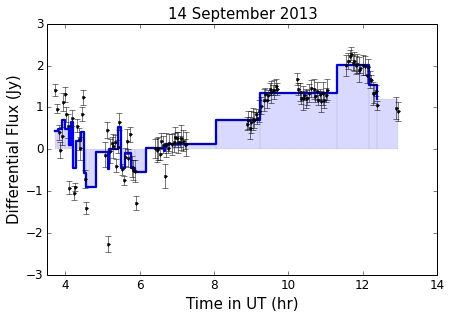}\par\caption{14 September 2013}
    \includegraphics[width=0.9\linewidth]{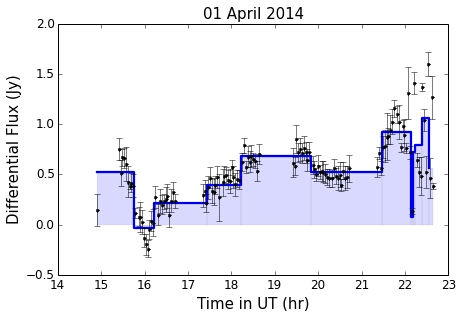}\par\caption{01 April 2014}
  \end{multicols}

\end{figure*}

\begin{figure*}
  \begin{multicols}{2}
    \includegraphics[width=0.9\linewidth]{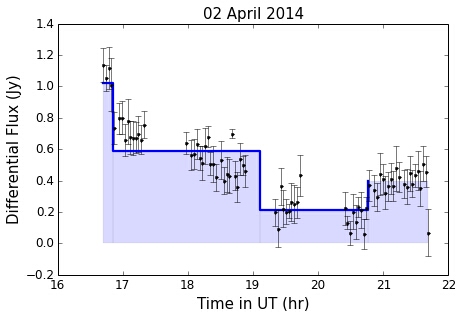}\par\caption{02 April 2014}
    \includegraphics[width=0.9\linewidth]{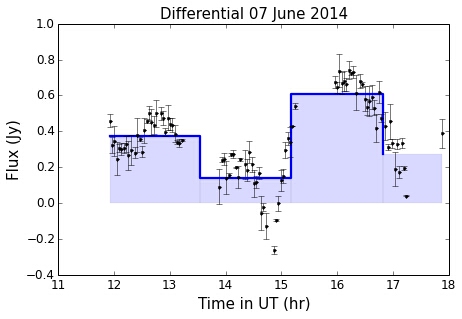}\par\caption{07 June 2014}
  \end{multicols}
  \begin{multicols}{2}
    \includegraphics[width=0.9\linewidth]{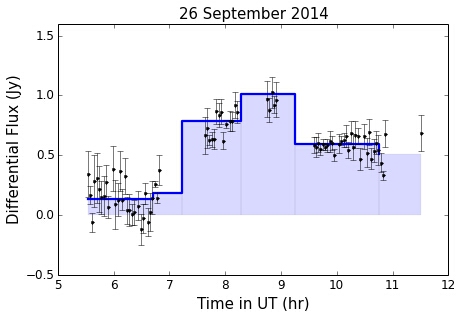}\par\caption{26 September 2014}
    \includegraphics[width=0.9\linewidth]{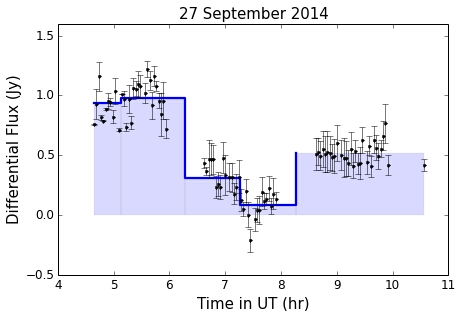}\par\caption{27 September 2014}
  \end{multicols}

\end{figure*}

%

\bsp

\label{lastpage}

\end{document}